# Graphene-Based Liquid Crystal Device

*Peter Blake[1], Paul D. Brimicombe[2], Rahul R. Nair[2], Tim J. Booth[3], Da Jiang[4], Fred Schedin[4], Leonid A. Ponomarenko[2], Sergey V. Morozov[5], Helen F. Gleeson[2], Ernie W. Hill[1], Andre K. Geim[4], Kostya S. Novoselov[2*]*

[1] School of Computer Science, University of Manchester, Manchester, M13 9PL, UK

[2] School of Physics & Astronomy, University of Manchester, Manchester, M13 9PL, UK

[3] Graphene Industries, Manchester, UK

[4] Centre for Mesoscience and Nanotechnology, University of Manchester, Manchester M13 9PL, UK

[5] Institute for Microelectronics Technology, 142432 Chernogolovka, Russia

kostya@manchester.ac.uk

Graphene is only one atom thick, optically transparent, chemically inert and an excellent conductor. These properties seem to make this material an excellent candidate for applications in various photonic devices that require conducting but transparent thin films. In this letter we demonstrate liquid crystal devices with electrodes made of graphene which show excellent performance with a high contrast ratio. We also discuss the advantages of graphene compared to conventionally-used metal oxides in terms of low resistivity, high transparency and chemical stability.

Graphene is the first example of truly two-dimensional materials[1]. Only one atom thick, it demonstrates high crystallographic quality[2] and ballistic electron transport on the micrometer scale. Such unique properties make it a realistic candidate for a number of electronic applications. In particular, graphene is an attractive material for optoelectronic devices, in which its high optical transmittance, low resistivity, high chemical stability and mechanical strength seems to make it an ideal optically-transparent conductor.

Transparent conductors are an essential part of many optical devices. Traditionally, thin metallic or metal-oxide films are used for these purposes (for a review see[3]). At the same time there is a constant search for new types of conductive thin films, as existing technologies are often complicated (e.g. thin metallic films require anti-reflection coating) and expensive (often using noble or rare metals). Furthermore, many of the widely used metal oxides exhibit nonuniform absorption across the visible spectrum[4] and are chemically unstable (for instance the commonly used Indium Tin Oxide (ITO, $In_2O_3$:Sn) is known to inject oxygen[5] and indium[6] ions into the active media of a device). Recently carbon nanotube films have been produced[7] and used as an alternative transparent conductor in various photonic devices including electric field-activated optical modulators, organic solar cells[8] and liquid crystal displays[9]. The experimental discovery of graphene[10] brought a new alternative to the ubiquitous ITO. The optical properties of this material are now being widely tested[11,12,13,14,15], and graphene films have recently been used as transparent electrodes for solar cells[16].

In this letter we report on the use of graphene as a transparent conductive coating for photonic devices and show that its high transparency and low resistivity make this two-dimensional crystal ideally suitable for electrodes in liquid crystal devices. We will also argue that graphene, being mechanically strong, chemically stable and inert, should improve the durability and simplify the technology of potential optoelectronic devices.

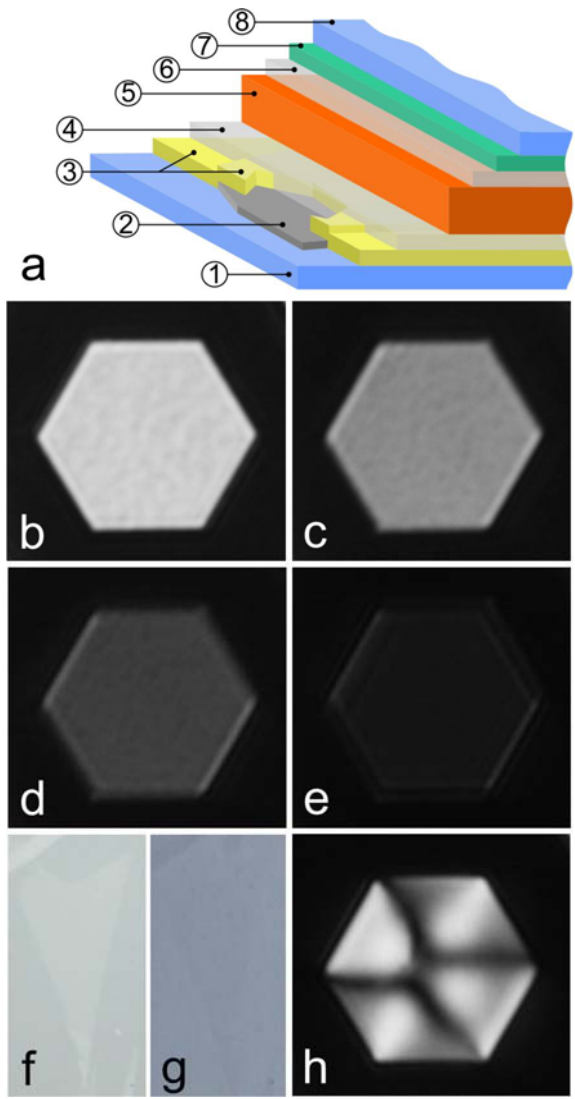

**Figure 1.** (a) Schematic diagram of our liquid crystal devices with typical layer thicknesses in brackets: 1 – glass (1mm); 2 – graphene; 3 – Cr/Au contact surrounding graphene flake (5nm Cr + 50nm Au); 4 – alignment layer (polyvinyl alcohol) (40 nm); 5 – liquid crystal (20μm); 6 – alignment layer (40 nm); 7 – ITO (150 nm); 8 – glass (1 mm). The graphene flake is surrounded by a non-transparent Cr/Au contact. (b-e) Optical micrographs of one of our liquid crystal device using green light (505 nm, FWHM 23 nm) with different voltages applied across the cell: (b) *V*=8 Vrms; (c) *V*=13 Vrms; (d) *V*= 22Vrms; (e) *V*=100 Vrms. Overall image width is 30μm. The central hexagonal window is covered by graphene, surrounded by the opaque Cr/Au electrode. (f) An optical micrograph (in reflection, using white light) of a graphene flake on the surface of a 1 mm thickness glass slide. The contrast is of the order of 6%. Overall image width is 10μm. (g)

The same image but in transmission. The flake is practically invisible. (h) Control device with no graphene in the opening of the Cr/Au contacts with *V*=100 Vrms applied across the cell. Since the electrode on the ITO-coated surface is continuous, there is a significant stray field within the window that distorts the liquid crystal structure, leading to the pattern shown.

Graphene flakes were prepared by micromechanical cleavage[10,17] on a glass microscope slide. They were first located using an optical microscope[18] (Figure 1f,g) and then further identified as monolayer graphene using Raman microscopy[19]. Thin (70 nm) chromium/gold contacts were then deposited around the flakes, so the graphene crystal was effectively covering a window in the metallization, Figure 1a,b (this geometry also eliminates stray electric fields from the edges of the electrode). A planar-aligned liquid crystal devices were then fabricated using such graphene-on-glass films as one of the transparent electrodes, Figure 1a. The other substrate was of a glass slide coated with conventional ITO. Both substrates were coated with a polyvinyl alcohol alignment layer which was subsequently baked and then unidirectionally rubbed (ITO-coated substrate only) in order to promote uniform alignment of the liquid crystal director. The device was then capillary-filled with nematic liquid crystal material E7 (Merck). Applying a voltage across the liquid crystal layer distorts the crystal alignment, changing the effective birefringence of the device and altering the transmitted light intensity[20]. A control sample, with an opening in the metallization not covered by graphene, was also prepared (Figure 1h). Note that although we will limit our consideration of graphene-based liquid crystal devices by those with planar untwisted nematic liquid crystals, this technology could equally be applied to any of the various nematic liquid crystal device types (e.g. twisted nematic[21], supertwisted nematic[22], in-plane switching[23] and vertically aligned nematic[24] devices) and also to ferroelectric and other liquid crystal devices that use smectic phases.

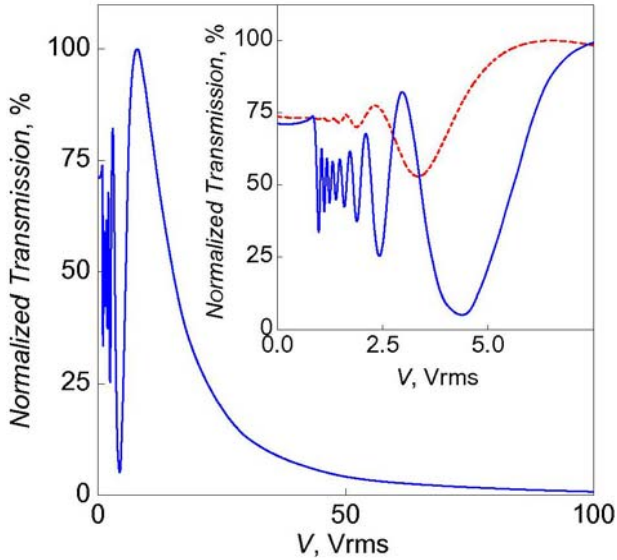

**Figure 2**. Light transmission through the liquid crystal device as a function of voltage applied across the cell, normalized to the maximum transmission. Inset: the same at low voltages. Solid blue curve: in green light, 505 nm, FWHM 23 nm; dashed red curve: in white light. The data taken in white light practically coincide with those in green light for voltages above 10Vrms and are omitted from the main panel for clarity. From the oscillatory behavior the thickness of the liquid crystal layer is estimated to be ~20 μm, assuming that the birefringence of E7 is 0.225.

An AC (square-wave) voltage was applied across the cells in order to reorient the liquid crystal director. The electro-optic properties were observed using an optical microscope with the device placed between crossed polarizers and the rubbing direction oriented 45° with respect to the polarizers. Above the expected threshold voltage of around 0.9 Vrms, a strong change in the transmission is observed (Figure 1b-e, 2) both in white and monochromatic light. The fact that the whole electrode area changes contrast uniformly suggests that the electric field is applied uniformly through the area of graphene and that the graphene has no negative effect on the liquid crystal alignment. The contrast ratio (between maximum transmission and the transmission when 100 Vrms is applied across the cell) is better than 100 under illumination using white light, which is very good for this type of cell and demonstrates that graphene could indeed be used effectively as a transparent electrode for liquid crystal displays. No significant changes in transmission were observed for the control sample, with only edge effects appearing due to the finite thickness of the cell, Figure 1h.

We will now assess the quality of our liquid crystal devices, concentrating on such important issues as the transparency of graphene, its resistivity and chemical stability. Light absorption by graphene is presented on Figure 3(right inset) as a function of the number of layers. Each layer of graphene absorbs about 2%, which is significantly lower than that of conventionally used ITO (15-18% ). Such high transmittance is explained by a low electronic density of states in graphene[14,15].

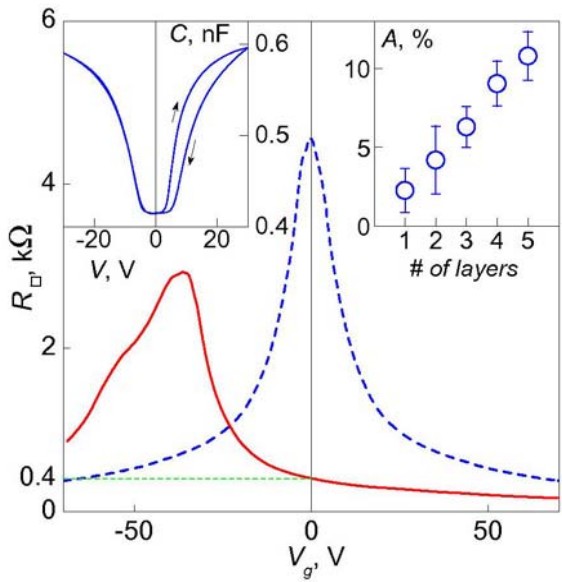

**Figure 3.** Sheet resistance of a graphene device as a function of gate voltage with (solid red curve) and without (dashed blue curve) a layer of polyvinyl alcohol on top. Polyvinyl alcohol provides n-type doping, shifting the curve to negative gate voltages. The sheet resistance at zero gate voltage is ∼400 Ω. Left inset: capacitance of one of our liquid crystal devices as a function of voltage applied. Right inset: light absorption of free hanging graphene of different thicknesses.

The sheet resistance of undoped graphene is of the order of 6kΩ (one conductivity quantum per species of charge carriers). However, it can be reduced down to 50Ω by chemical doping[10,25], and even unintentional doping (due to molecules absorbed from the surrounding atmosphere, e.g. water) can be of the order of $10^{12}$ cm$^{-2}$). In liquid crystal devices an electrode is usually in direct contact with an alignment layer (in our case polyvinyl alcohol). We have tested the doping of graphene with polyvinyl alcohol, by preparing a standard graphene device on a 300nm SiO$_2$/Si wafer and measuring its gate response with and without a layer of polyvinyl alcohol on top of graphene (Figure 3). The introduction of a layer of polyvinyl alcohol produces n-type doping of about $3\times10^{12}$ cm$^{-2}$. For this particular sample it resulted in a drop in the sheet resistance down to 400Ω, which is an impressive result for a conductive coating with optical transmission of about 98%. It is difficult to compare this result to ITO, as the resistance of In$_2$O$_3$:Sn films diverges strongly (in the order of tens of kΩ) when trying to obtain optical transmittance above 95%. ITO films with 95% transmittance demonstrate comparable sheet resistances of a few hundred Ohms, dropping to tens of Ohms at an optical transmittance of about 90%[26]. Similar or even lower resistances can be achieved for graphene by a variety of means: increasing the number of layers[27], intentional doping, or by using samples with higher mobility[28,29].

An important issue for most ITO-based liquid crystal devices and other photonic devices is the chemical stability of the metal-oxide and the diffusion of ions into the active media. Such processes deteriorate the active media (for example via oxidation if oxygen is injected) and can lead to break-down at lower voltages. Furthermore, in liquid crystal displays the injected ions get trapped at the alignment layer, thus screening the applied electric field. This leads to the so-called image sticking problem[30] which is usually avoided by driving the liquid crystal cells with alternating voltage.

One can generally expect that such issues can be avoided when using graphene, where its chemical stability should minimize the level of ion diffusion. To check this we have measured the capacitance of one of our liquid crystal devices, which has one electrode made of graphene and the other from ITO, when applying DC voltages of different polarities (Figure 3 left inset, here positive voltage corresponds to higher potential on the ITO electrode). There is clearly a highly hysteretic response when applying positive biases, but no hysteresis has been observed at the opposite polarity. We attribute this observation to positive indium ions drifting into the liquid crystal from the ITO electrode, whereas no ions are injected from the graphene electrode. Similar liquid crystal devices constructed using ITO electrodes on both substrates produce the hysteretic response for both polarities.

Although it is important to demonstrate the possibility and advantages of using graphene as a transparent conductive coating, the feasibility of its mass production is essential when considering realistic applications. No industrial technology can rely on the micromechanical cleavage technique that allows only minute quantities of graphene and, although sufficient for fundamental research and proof-of-concept devices, is unlikely to become commercially viable. Recently, large-area conductive films have been demonstrated by using chemical exfoliation of graphite oxide and then reducing it to graphene[16,31,32]. This could lead to a viable way of making thin graphene-based films with properties similar to those discussed earlier and using them for various photonic devices. However, so far this technique has not demonstrated the ability to fully recover the excellent conductive properties of graphene to recover[33]. We propose an alternative approach. It involves making a graphene suspension by direct chemical exfoliation of graphite (rather than graphite oxide), which is subsequently used to obtain transparent conductive films on top of glass by spin- or spray-coating.

Crystals of natural graphite (Branwell Graphite Ltd.) were exfoliated by sonication in dimethylformamide (DMF) for over 3 hours. DMF "dissolves" graphite surprisingly well, and the procedure resulted in a suspension of thin graphitic platelets with large proportion of monolayer graphene flakes, DMF also wets the flakes preventing them from conglomerating[34]. The suspension was then centrifuged at 13,000 rpm for 10 minutes to remove thick flakes. The remaining suspension, consisting mostly of graphene and few-layer graphite flakes of sub-micrometer size, was spray-deposited onto a preheated glass slide (Figure 4a,b) which yielded thin (~1.5nm) films over centimeter sized areas. These films were then annealed for 2 hours in argon(90%)/hydrogen(10%) atmosphere at 250°C. The transparency of such graphitic layers was approximately 90% (Figure 4b).

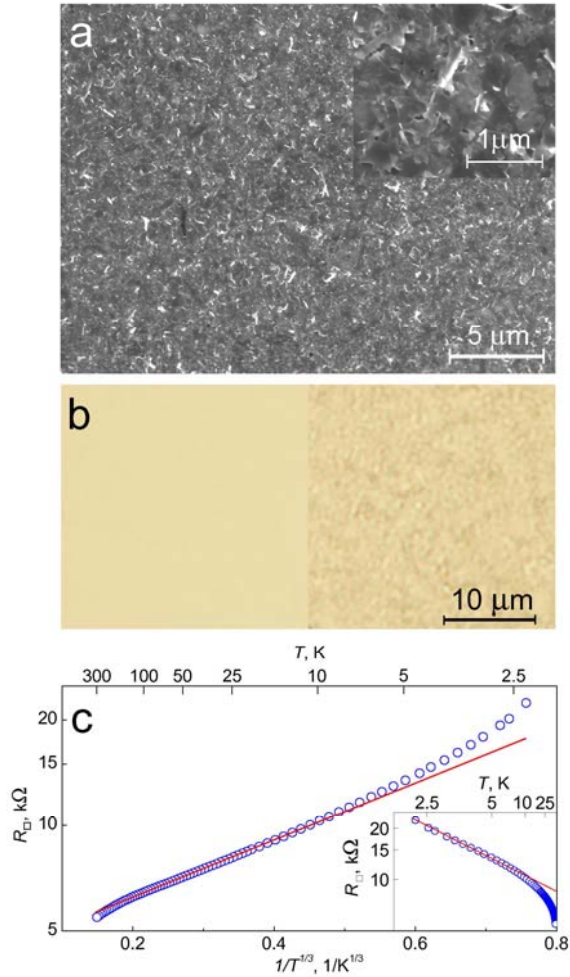

**Figure 4.** (a) Scanning electron micrograph of a thin graphitic film obtained by chemical exfoliation and spray-coating. Inset shows the same area under higher magnification. (b) Light transmission through an original glass slide (left) and the one covered with the graphitic film (right). (c) Temperature dependence of the film's sheet resistance ($R \sim \exp(T_0/T^{1/3})$ behavior is observed at $T>10$K, where $T_0$ is a constant). Inset: the same data but for the low temperature interval ($R \sim \exp(\Delta/T)$ behavior is observed at T<10K, where $\Delta$ is a constant). The red lines are guides for the eye.

In order to measure resistivity of our films, a mesa structure in the shape of the Hall bar with typical dimensions of 1mm was prepared, and the four-probe resistance was measured as a function of temperature (Figure 4c). The high temperature region (above 10 K) is well described by $\exp(T_0/T^{1/3})$ dependence, characteristic for variable range hopping in two dimensions[35]. The room temperature sheet resistance is of the order of 5 kΩ, which is already acceptable for some applications[3,16], and can be decreased further by increasing the film thickness. Resistance at low temperatures deviates from the variable-range-hopping dependence but can be described by the simple activation dependence $\exp(-\Delta/T)$ (see inset in Figure 4c). We attribute this low-temperature behaviour to weak tunneling-like coupling between different flakes, possibly due to contamination with organic (DMF) residues. This indicates some potential for improvements as better cleaning and annealing procedures can potentially improve coupling between graphene crystallites and decrease the film resistance further.

To conclude, high optical transparency, low resistivity and high chemical stability of graphene makes it an excellent choice for transparent electrodes in various optoelectronic devices. Furthermore, there are already several technologies that potentially allow mass production of thin graphene-based transparent conductors (besides the chemical exfoliation of graphite described in the present letter, one can also think of epitaxial growth of graphene on top of a metal surface, followed by a transfer of such a layer onto a transparent substrate). These techniques are capable of producing continuous graphene films of thickness below 5 monolayers, which is required for realistic applications.

The authors are grateful to EPSRC for financial support. AKG and KSN also acknowledge support from the Royal Society, UK.